# Phase randomness in a gain-switched semiconductor laser: stochastic differential equation analysis

Roman Shakhovoy[1,2,3*], Alexander Tumachek[2], Natalia Andronova[2], Yury Mironov[2], and Yury Kurochkin[1,3,4]

[1]*QRate, 100 Novaya str., Skolkovo, Russian Federation*
[2]*Moscow Technical University of Communications and Informatics, 8a Aviamotornaya str., Moscow, Russian Federation*
[3]*NTI Center for Quantum Communications, National University of Science and Technology MISiS, 4 Leninsky prospekt, Moscow, Russian Federation*
[4]*Russian Quantum Center, 45 Skolkovskoye shosse, Moscow, Russian Federation*
*[*]r.shakhovoy@goqrate.com*

**Abstract**

We performed theoretical analysis of the phase randomness in a gain-switched semiconductor laser in the context of its application as a quantum entropy source. Numerical simulations demonstrate that phase diffusion r.m.s. exhibits non-linear dependence on the bias current, which could be of significant practical importance, particularly, in application to high-speed optical quantum random number generators. It is shown that phase diffusion between laser pulses cannot always be assumed to exhibit required efficiency, particularly, at high pulse repetition rates. It was also revealed that the gain saturation significantly affects the r.m.s. value of the phase diffusion and, in essence, determines the degree of non-linearity of its dependence on the pump current.

**1. Introduction**

Phase randomness between pulses of a gain-switched semiconductor laser is an essential ingredient of quantum key distribution (QKD) systems and quantum random number generators (QRNGs). Inasmuch as amplified spontaneous emission (ASE) dominates below laser threshold [1, 2], phase correlations of the electromagnetic field should be destroyed very quickly between laser pulses under the gain switching. Therefore, most authors assume that pulses from a gain-switched laser have no phase relationship to each other. Indeed, the security analysis of QKD protocols, particularly decoy-BB84 protocol [3-5], implicitly assumes that the laser emits light in a mixture of coherent states with uniformly distributed phases. Similar assumption is usually made when considering laser pulse interference as a source of quantum entropy for some optical QRNGs [6-13]. In real experiments, however, phase correlations may still occur and may thus lead to problems in terms of information security. So, phase diffusion between laser pulses should be treated more carefully.

The main reason of correlations between phases of pulses emitted by a gain-switched semiconductor laser is an insufficient delay between subsequent pulses, during which the phase does not have enough time to "diffuse". Another reason is a high value of the bias current, which does not allow the carrier number to get well below the laser threshold between pulses. It was estimated in [6] that enough randomness could be achieved with the laser pulse repetition rate up to 20 GHz in assumption that attenuation between pulses reaches 100 dB. The authors demonstrated later an optical QRNG with the distributed feedback (DFB) gain-switched laser operating at pulse repetition rate of 5.825 GHz [7]. As a criterion of the phase randomness, they used the variance of phase fluctuations, which was estimated from the relation derived by C. Henry [14] for the phase diffusion *above* threshold. Although such an approach can be used, e.g., to follow the qualitative dependence of the phase diffusion on the pump current, it is not fully faithful for quantitative estimation, since, as we noted above, phase randomization in a gain-switched laser occurs due to the phase diffusion *below* threshold. Another approach has been developed in [8], where authors used visibility of interference fringes as a criterion of phase randomness. Such an approach provides a useful experimental probe to estimate the effectiveness of the phase diffusion; however, it does not allow distinguishing classical and pure quantum contributions to phase fluctuations, which is important for QRNG.

In principle, any experimental measurement of the phase diffusion will be affected by the presence of the classical (e.g. thermal) noise in electronic and fiber-optic components. Therefore, it would be extremely useful to have in addition a rigorous theoretical model, which will allow analyzing dependence of the phase diffusion on laser parameters as well as evaluating pure quantum component of the phase noise. The most natural choice for such a model is a system of semiconductor laser rate equations with stochastic (Langevin) terms [1, 14-17]. In the present study, we use Langevin equations to analyze the phase diffusion in a gain-switched semiconductor laser. Our analysis provides a clear criterion for the phase diffusion efficiency at any pulse repetition rate and any value of the pump current.

**2. Stochastic differential equations**

The system of rate equations for a single-mode semiconductor laser can be written as follows [1, 14-17]:

$$dQ = (G-1)\frac{Q}{\tau_{ph}}dt + C_{sp}R_{sp}dt + F_Q dt,$$

$$d\varphi = \frac{\alpha}{2\tau_{ph}}(G_L - 1)dt + F_\varphi dt, \qquad (1)$$

$$dN = \frac{I}{e}dt - Rdt - \frac{QG}{\Gamma\tau_{ph}}dt + F_N dt.$$

Here $\varphi$ is the phase of the electric field, $N$ is the carrier number, $I$ is the pump current, $e$ is the absolute value of the electron charge, $R$ is the rate of spontaneous recombination (including both radiative and non-radiative), $R_{sp}$ is the rate of radiative spontaneous recombination, the factor $C_{sp}$ corresponds to the fraction of spontaneously emitted photons that end up in the active mode, $\alpha$ is the linewidth enhancement factor (the Henry factor [14]), and the dimensionless linear gain $G_L$ is defined by $G_L = (N - N_{tr})/(N_{th} - N_{tr})$, where $N_{tr}$ and $N_{th}$ are the carrier numbers at transparency and threshold, respectively. Onwards, $Q$ is the intensity of the normalized electric field, $Q \equiv |E|^2$, defined such that its average corresponds to the average photon number $\langle n_{ph} \rangle$, and $\tau_{ph}$ is a characteristic time of the field intensity decay in a laser cavity, which is usually referred to as the photon lifetime. Note that strictly speaking the quantity $Q$ cannot be treated as the photon number and should be rather referred to as a normalized intensity. This feature is not relevant when considering rate equations for averaged quantities, since $\langle Q \rangle = \langle n_{ph} \rangle$; however, it becomes relevant when considering stochastic rate equations, since quantities $n_{ph}$ and $Q$ have different distributions and diffusion coefficients [15]. $Q$ is related to the output power $P$ by $P = Q(\varepsilon\hbar\omega_0/2\Gamma\tau_{ph})$, where $\hbar\omega_0$ is the photon energy ($\omega_0$ is the carrier frequency), $\varepsilon$ is the differential quantum output, $\Gamma$ is the confinement factor, and the factor $1/2$ takes into account the fact that the output power is generally measured from only one facet. Note also that we included into Eq. (1) the gain saturation [18] by using the following relation: $G = G_L(1 - \chi P)$, where $\chi$ is the gain compression factor [1]. Finally, Langevin forces $F_Q$, $F_\varphi$, $F_N$ in Eq. (1) drive fluctuations of $Q$, $\varphi$, and $N$, respectively, and obey the relation $\langle F_a(t)F_b(t-\tau)\rangle = 2D_{ab}\delta(\tau)$, where $\delta(\tau)$ is the delta-function and $D_{ab}$ are diffusion coefficients (see the supplementary material for details). We will assume further for simplicity that radiative spontaneous recombination prevails over the non-radiative one, such that one can write $R_{sp} \approx R = N/\tau_e$, where $\tau_e$ is the effective lifetime of the electron.

Note that diffusion coefficients characterize only the statistical distribution of the noise, but they do not describe the effects of the noise on the laser dynamics. Therefore, integration of stochastic rate equations (1) requires an explicit form of stochastic terms. Despite a well-developed theory of Langevin equations, ambiguity is still present in the literature concerning an explicit form of Langevin forces in Eq. (1) [19-21]. Recently [21], Langevin equations for a semiconductor laser were reformulated in a mathematical framework of the Itô calculus [22], where the system containing Langevin noise terms was rigorously interpreted by defining Wiener processes that drive the noise. Within this formalism, stochastic differential equations can be written in the following form ready for numerical integration (see derivation in the supplementary material):

$$Q_{n+1} = Q_n + \left(\frac{N_n - N_0}{N_{th} - N_0}(1 - \chi_Q Q_n) - 1\right)\frac{Q_n}{\tau_{ph}}\Delta + C_{sp}\frac{N_n}{\tau_e}$$

$$+ 2\sqrt{\frac{C_{sp}\bar{N}_n\bar{Q}_n}{2\tau_e}}\left(\xi_n^A \cos\bar{\varphi}_n + \xi_n^B \sin\bar{\varphi}_n\right)\sqrt{\Delta},$$

$$\varphi_{n+1} = \varphi_n + \frac{\alpha}{2\tau_{ph}}\left(\frac{N_n - N_0}{N_{th} - N_0} - 1\right)\Delta$$

$$+ \sqrt{\frac{C_{sp}\bar{N}_n}{2\tau_e\bar{Q}_n}}\left(\xi_n^B \cos\bar{\varphi}_n - \xi_n^A \sin\bar{\varphi}_n\right)\sqrt{\Delta}, \qquad (2)$$

$$N_{n+1} = N_n + \frac{I_n}{e}\Delta - \frac{N_n}{\tau_e}\Delta - \frac{Q_n}{\Gamma\tau_{ph}}\frac{N_n - N_0}{N_{th} - N_0}(1 - \chi_Q Q_n)\Delta$$

$$- 2\sqrt{\frac{C_{sp}\bar{N}_n\bar{Q}_n}{2\tau_e}}\left(\xi_n^A \cos\bar{\varphi}_n + \xi_n^B \sin\bar{\varphi}_n\right)\sqrt{\Delta}$$

$$+ \sqrt{\frac{2\bar{N}_n}{\tau_e}}\xi_n^C \sqrt{\Delta},$$

where $\Delta = t_{n+1} - t_n$ is the time increment, $\chi_Q = \chi\left(\varepsilon\hbar\omega_0/2\Gamma\tau_{ph}\right)$ is the dimensionless gain compression factor, $\xi^A$, $\xi^B$, and $\xi^C$ are independent discrete random variables with standard normal distribution, and, finally, $\bar{N}$, $\bar{Q}$, and $\bar{\varphi}$ can be found from the same system of rate equations without stochastic terms. In a sense, $\bar{N}(t)$, $\bar{Q}(t)$ and $\bar{\varphi}(t)$ in Eq. (2) are predefined functions of time, just like the pump current $I(t)$; therefore, to make this system self-consistent, one should use the following initial values: $N_0 = \bar{N}_0$, $Q_0 = \bar{Q}_0$, and $\varphi_0 = \bar{\varphi}_0$. Note also that direct implementation of the Euler-Mariama scheme may lead to an unphysical result, namely to negative values of $N$ and $Q$; therefore, Eq. (2) should be solved with constraint that $N$ and $Q$ are non-negative. It can be performed, e.g., by equating $Q_n$ to zero, when it becomes negative (and the same for $N_n$).

### 3. Monte-Carlo simulations

Using Eq. (2) we performed Monte-Carlo simulations of the phase diffusion between adjacent optical pulses of a gain-switched laser for different values of the pulse repetition rate $f_p$: 2.5 GHz, 5 GHz, and 10 GHz. At each frequency, the pump current was assumed to have a form of a square wave, $I(t) = I_b + I_p(t)$, where $I_b$ is the bias current and $I_p(t)$ changed abruptly from 0 to $I_p$. We performed integration of Eq. (2) from 0 to $T_p$, i.e. the phase was allowed to diffuse during the time $T_p = 1/f_p$. After each such integration, we obtained a value $\varphi(T_p)$ of a random phase, which exhibited Gaussian distribution, whose r.m.s we denote as $\sigma_\varphi(T_p) \equiv \sigma_\varphi$. 50000 iterations (random values of $\varphi(T_p)$) were found to be enough to find a value of $\sigma_\varphi$ for given values of $I_b$, $I_p$ and a given set of laser parameters. In Fig. 1, we presented the dependence of $\sigma_\varphi$ on the bias current $I_b$ at different values of $f_p$. At each pulse repetition rate, we calculated several curves corresponding to different values of $I_p$. For each $I_p$, we chose the range of the bias current variation from the value $I_b^{min}$ corresponding to stable pulsation at a given $I_p$ up to the threshold value defined by $I_{th} = N_{th}e/\tau_e$. The gain compression factor $\chi$ was put to 20 W$^{-1}$. Other laser parameters used in simulations are listed in Table I. In order to achieve high performance and reduce the computation time, we performed computations on a Compute Unified Device Architecture (CUDA) platform with the Nvidia video card equipped by a CUDA-powered graphics processing unit.

### 4. Results and discussion

One can see from Fig. 1 that the phase diffusion r.m.s. increases towards lower values of the bias current. This reflects the fact that the number of carriers decreases faster at a lower pump current between laser pulses, which leads to a faster decrease of the field intensity inside the laser cavity and, consequently, to a faster decoherence due to spontaneous emission. An interesting feature here is a non-monotonic behavior of $\sigma_\varphi$ curves, which exhibit some kind of "damped oscillations" (see Fig. 1). This is an interesting result, which could be of significant practical importance, particularly, in the context of optical QRNGs.

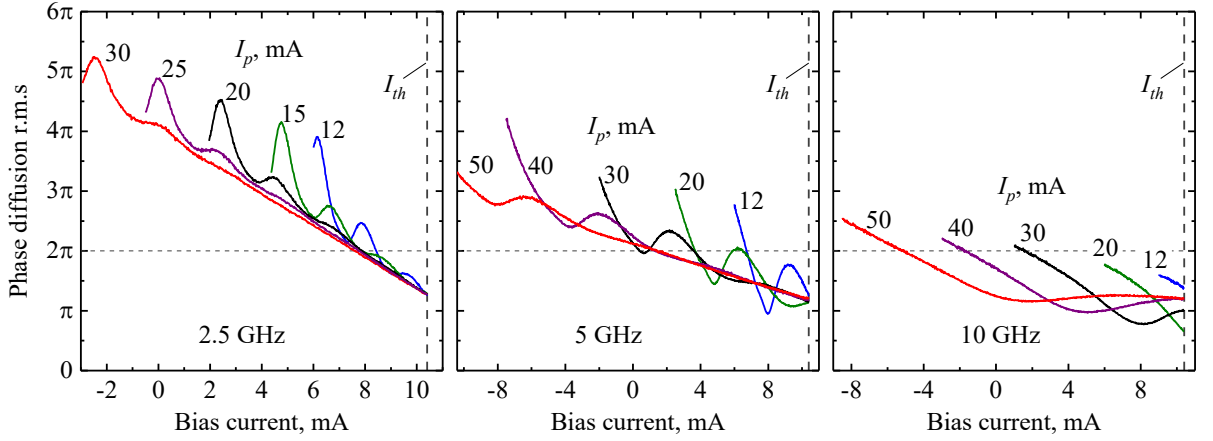

Fig. 1. Phase diffusion r.m.s. as a function of the bias current at different peak-to-peak values of the modulation current $I_p$ and at different pulse repetition rates. Vertical dashed lines denote the threshold current value. Corresponding laser parameters are listed in Table I.

In most applications, phase randomness between individual pulses is transformed into random intensity variations with the use of unbalanced interferometer (e.g., Mach-Zehnder or Michelson). The delay line of the interferometer is selected such that the corresponding delay time is a multiple of the pulse repetition period, so one measures the interference of pulses emitted by the laser at different moments of time. In this case, the random optical signal is related to the phase difference $\Delta\Phi$ between interfering pulses via $\cos(\Delta\Phi)$. If $\Delta\Phi$ is a Gaussian random variable with the r.m.s. equal to $\sigma_{\Delta\Phi}$, probability distribution of $\Delta\Phi$ may be substituted by the uniform distribution in the interval $[0,\pi)$, if $\sigma_{\Delta\Phi} > 2\pi$ [13]. This inequality can be considered as a condition of an absolute randomness in a sense that when the r.m.s. falls below this value, fluctuations of $\Delta\Phi$ are not random enough to be used as a purely quantum entropy source and one should take into account the contribution of classical noise into $\Delta\Phi$ fluctuations [13].

When considering phases of interfering laser pulses as two independent Gaussian random variables, each with the r.m.s. equal to $\sigma_\varphi$ (such an approach is sometimes convenient for simulations and was used in [13]), $\sigma_{\Delta\Phi}$ may be set equal to $\sigma_\varphi\sqrt{2}$. However, in a real experiment on the pulse interference, $\Delta\Phi$ corresponds to the phase incursion itself over time $T_p$: $\Delta\Phi = \varphi(T_p)$; therefore, we should put $\sigma_{\Delta\Phi} = \sigma_\varphi(T_p)$, and the inequality $\sigma_{\Delta\Phi} > 2\pi$ is met when $\sigma_\varphi > 2\pi$. It should be noted here that the results obtained in Fig. 1 remain practically unchanged even in the absence of fluctuations in the carrier number. In other words, the phase diffusion curves presented in the figure can be considered to be related to purely quantum noise. So, the point on the curve satisfying the condition $\sigma_\varphi > 2\pi$ corresponds to the parameters, at which the gain-switched laser is a true quantum entropy source.

TABLE I. Laser Parameters Used in Simulations

| Parameter | Value |
|---|---|
| Photon lifetime $\tau_{ph}$, ps | 1.0 |
| Electron lifetime $\tau_e$, ns | 1.0 |
| Quantum differential output $\varepsilon$ | 0.3 |
| Transparency carrier number $N_{tr}$ | $6.0 \times 10^7$ |
| Threshold carrier number $N_{th}$ | $6.5 \times 10^7$ |
| Spontaneous emission coupling factor $C_{sp}$ | $10^{-5}$ |
| Confinement factor $\Gamma$ | 0.12 |
| Linewidth enhancement factor $\alpha$ | 6 |
| Central lasing frequency $\omega_0/2\pi$, THz | 193.548 |

The boundary $\sigma_\varphi = 2\pi$ is shown in Fig. 1 by the horizontal dashed lines. According to the figure, the requirement $\sigma_\varphi > 2\pi$ is satisfied at $f_p = 2.5$ GHz for any appropriate value of $I_p$ when $I_b < 7$ mA. However, at higher values of the bias current, this requirement can be achieved only at $I_p < 12$ mA. This result may seem counterintuitive, since at first glance the higher peak-to-peak value of the modulation current (at a given value of the bias current) should provide more efficient phase diffusion. Such a result is caused by non-monotonic (oscillating) behavior of the $\sigma_\varphi(I_b)$ curves. Due to these oscillations, the phase diffusion may be more efficient when employing a smaller peak-to-peak value of the modulation current $I_p$, if $I_b$ approaches threshold value. This result can be used in the form of the following practical recommendation: in order to achieve the most efficient phase diffusion at a given $I_p$ value, one should set the bias current $I_b$ close to $I_b^{min}$, the value corresponding to stable pulsation.

Obviously, the requirement $\sigma_\varphi > 2\pi$ is more difficult to satisfy at higher pulse repetition rates. One can see, that already at $f_p = 5$ GHz the phase diffusion r.m.s. is guaranteed to be greater than $2\pi$ only when $I_b < 0$, i.e. for the reverse-biased diode, although the positive bias current could also be used with the carefully chosen $I_p$ value. Finally, at $f_p = 10$ GHz, the condition $\sigma_\varphi > 2\pi$ is hardly possible to satisfy with positive bias current, and only reverse-biased diodes can be used at such high repetition rates, which, in turn, forces the use of quite large peak-to-peak values of the modulation current.

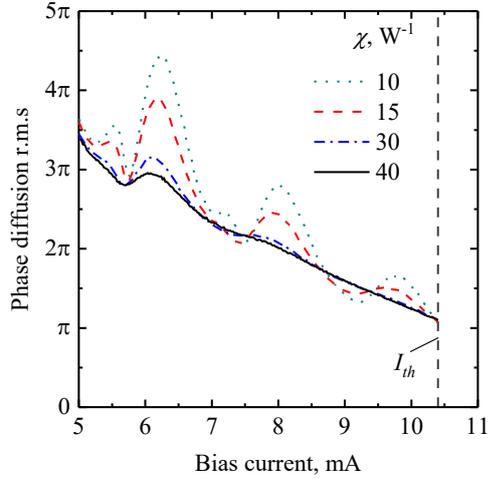

Fig. 2. Phase diffusion r.m.s. as a function of the bias current at four different values of the gain compression factor $\chi$. The peak-to-peak value of the modulation current and the pulse repetition rate were $I_p = 12$ mA and $f_p = 2.5$ GHz, respectively. Other parameters are listed in Table I.

The nature of the revealed oscillations in the $\sigma_\varphi(I_b)$ curves can be understood by referring to the work of C. Henry [23], where the author provides an explanation for the additional peaks appearing in a spectrum of a semiconductor laser. The satellite peaks were related to the time dependence of the mean square phase change $\langle \Delta\varphi^2 \rangle$, which, in addition to a linear dependence, exhibits damped periodical variations caused by relaxation oscillations. It was shown that

$$\langle \Delta\varphi^2 \rangle \propto \exp(-\gamma_d t)\cos(\omega_r t - 3\delta) \tag{3}$$

where $\gamma_d$ and $\omega_r$ are the damping rate and the angular frequency of relaxation oscillations, respectively, and $\cos(\delta) = \omega_r / \sqrt{\omega_r^2 + \gamma_d^2}$. Obviously, the quantity $\sigma_\varphi$ we have considered so far corresponds to $\langle \Delta\varphi^2 \rangle^{1/2}$ and should thus exhibit similar features. Note, however, that the time is fixed in our case to $t = T_p$; therefore, oscillations appearing in $\sigma_\varphi(I_b)$ are not related to the change of $t$. One can show using the small-signal analysis that $\omega_r$ can be approximated by the following relation [1]:

$$\omega_r \approx \sqrt{\frac{I_b + I_p - I_{th}}{I_{th} - I_{tr}} \frac{1}{\tau_{ph}\tau_e}}, \tag{4}$$

where we put $I_{tr} = N_{tr}e/\tau_e$. So, the damped oscillations of $\sigma_\varphi$ should be related to the change of the relaxation oscillation frequency $\omega_r$ governed by the change of the bias current $I_b$. It should be also noted that analytical expression for the time dependence of $\langle \Delta\varphi^2 \rangle$ derived by C. Henry is valid only for the steady state laser emission above threshold; therefore, such explanation is just an intuitive picture, which can be used to understand the origin of the $\sigma_\varphi$ oscillations, whereas quantitative estimations should be performed only via numerical solution of Eq. (2).

One can also see from Eq. (3) that the amplitude of $\sigma_\varphi(I_b)$ variations should depend on $\gamma_d$. To follow this dependence, we performed simulations of the phase diffusion r.m.s. at $f_p = 2.5$ GHz and $I_p = 12$ mA with four different values of the gain compression factor $\chi$ (see Fig. 2). Indeed, the damping rate of relaxation oscillations can be written as follows [23]: $\gamma_d = \gamma'_d + \chi P/\tau_{ph}$ ($\gamma'_d$ is the part of $\gamma_d$, which does not depend on $\chi$), so, $\gamma_d$ increases when increasing $\chi$, which is clearly seen from Fig. 2. Of course, one may expect that similar dependences will take place also at steady-state emission above threshold (see the supplementary material).

**5. Conclusion**

In conclusion, performed simulations demonstrate that the phase diffusion between the pulses of a gain-switched laser exhibits non-linear dependence on the bias current. Obtained $\sigma_\varphi(I_b)$ curves were shown to be related with relaxation oscillations and to have the same origin as the satellite peaks appearing in the spectrum of a semiconductor laser. As a general rule, we may conclude that in order to achieve more efficient phase randomization a smaller bias current is preferable. Another important result, which may be also useful from a practical point of view, is that at a bias current value close to the threshold the phase diffusion could be much more efficient at smaller values of the modulation current $I_p$. We also revealed that the gain saturation significantly affects the r.m.s. value of the phase diffusion and, in essence, determines the degree of non-linearity of its dependence on the pump current.

**Acknowledgments**
The authors acknowledge support from the Russian Science Foundation (Grant No. 17-71-20146)

SUPPLEMENTARY MATERIAL
for
**Phase randomness in a gain-switched semiconductor laser: stochastic differential equation analysis**
Roman Shakhovoy[1,2,3*], Aslexander Tumachek[2], Natalia Andronova[2], Yury Mironov[2], and Yury Kurochkin[1,3,4]

[1]*QRate, 100 Novaya str., Skolkovo, Russian Federation*
[2]*Moscow Technical University of Communications and Informatics, 8a Aviamotornaya str., Moscow, Russian Federation*
[3]*NTI Center for Quantum Communications, National University of Science and Technology MISiS, 4 Leninsky prospekt, Moscow, Russian Federation*
[4]*Russian Quantum Center, 45 Skolkovskoye shosse, Moscow, Russian Federation*
*\*r.shakhovoy@goqrate.com*

### *1. Derivation of stochastic differential equations*

To derive stochastic differential equations (SDEs), let us start from the rate equation for the slowly varying electric field amplitude $E$, which can be written in the following symbolic differential form [1, 2]:

$$dE = \frac{1}{2\tau_{ph}}(1+i\alpha)(G_L - 1)Edt + F_E(t)dt, \qquad (1)$$

where $\tau_{ph}$ is a characteristic time of the field intensity decay in a laser cavity, $\alpha$ is the linewidth enhancement factor (the Henry factor [3]), and the dimensionless gain $G_L$ is assumed to have linear dependence on the carrier number $N$: $G_L = (N - N_{tr})/(N_{th} - N_{tr})$, where $N_{tr}$ and $N_{th}$ are the carrier numbers at transparency and threshold, respectively. The normalized dimensionless complex field amplitude $E$ in Eq. (1) is defined such that its averaged absolute square, $\langle |E|^2 \rangle$, corresponds to the average photon number $\langle n_{ph} \rangle$. The time $\tau_{ph}$ can thus be treated as the photon lifetime.

The complex Langevin force $F_E(t) = F_E'(t) + iF_E''(t)$ in Eq. (1) is assumed to obey the relation [1]

$$\langle F_E(t) F_E^*(t-\tau) \rangle = C_{sp} \bar{R}_{sp}(t) \delta(\tau), \qquad (2)$$

where $\delta(\tau)$ is the delta-function, $\bar{R}_{sp}(t) = \bar{N}(t)/\tau_e$ is the rate of radiative spontaneous emission ($\tau_e$ is the carrier lifetime), and the phenomenological factor $C_{sp}$ corresponds to the fraction of spontaneously emitted photons that end up in the active mode under consideration. It is important to note that averaging in Eq. (2) is performed during the time short compared to the time needed for noticeable change of $Q$ and $N$. To indicate this fact explicitly, we use the bar under $R_{sp}$ in Eq. (2), which means that it does not fluctuate, but still depends on time $t$.

In the SDEs theory, Langevin forces are usually substituted by corresponding Wiener processes. Using Eq. (2), we can rewrite the diffusive term in Eq. (1) as $\sqrt{C_{sp}\bar{R}_{sp}/2}\left(\xi_t^A + i\xi_t^B\right)dt$, where $\xi_t^j$ are standard Gaussian random variables for each $t$, which obey the relations $\langle \xi_t^j \xi_{t-\tau}^j \rangle = \delta(\tau)$ and $\langle \xi_t^j \xi_{t-\tau}^k \rangle = \langle \xi_t^j \rangle \langle \xi_{t-\tau}^k \rangle = 0$ ($j \neq k$) (the subscript $t$ indicates here the time dependence). Transition to the Wiener process $W$ can be now performed with the following symbolic relation: $dW = \xi_t dt$, such that the last term in Eq. (1) is

$$F_E dt = \sqrt{\frac{C_{sp}\bar{R}_{sp}}{2}}(dW^A + idW^B). \qquad (3)$$

Equation (1) for the complex field $E$ contains actually two variables and is thus a vector valued SDE. One usually chooses the field phase $\varphi$ and the intensity $Q$ as such variables; however, it is convenient for further derivation to use the field $E$ and its complex conjugate $E^*$ instead. With such a choice, the vector valued SDE of interest can be written as

$$dX = a(X)dt + b(X)dW, \qquad (4)$$

where we introduced the vectors

$$X = \begin{pmatrix} E \\ E^* \end{pmatrix}, \quad W = \begin{pmatrix} W^A \\ W^B \end{pmatrix},$$

$$a(X) = \frac{G_L - 1}{2\tau_{ph}} \begin{pmatrix} (1+i\alpha)E \\ (1-i\alpha)E^* \end{pmatrix}, \qquad (5)$$

and the matrix

$$b(X) = \sqrt{\frac{C_{sp}\bar{R}_{sp}}{2}} \begin{pmatrix} 1 & i \\ 1 & -i \end{pmatrix}. \tag{6}$$

To derive an SDE for the normalized intensity $Q = EE^*$, we will need the Itô formula for the differential of the stochastic function $Y = U(X^1, X^2, ..., X^d, t)$. It is given by [4]

$$dY = \left\{ \frac{\partial U}{\partial t} + \sum_{k=1}^{d} a^k \frac{\partial U}{\partial X^k} + \frac{1}{2} \sum_{j=1}^{m} \sum_{i,k=1}^{d} b^{i,j} b^{k,j} \frac{\partial^2 U}{\partial X^i \partial X^k} \right\} dt$$
$$+ \sum_{j=1}^{m} \sum_{i=1}^{d} b^{i,j} \frac{\partial U}{\partial X^i} dW^j, \tag{7}$$

where $d$ and $m$ are dimensions of vectors $X$ and $W$, respectively (in our case, $m = d = 2$). For $Y = EE^*$ we obtain from Eq. (7)

$$dQ = (G_L - 1)\frac{Q}{\tau_{ph}} dt + C_{sp}\bar{R}_{sp} dt$$
$$+ 2\sqrt{\frac{C_{sp}\bar{R}_{sp}\bar{Q}}{2}} \left( \cos\bar{\varphi} dW^A + \sin\bar{\varphi} dW^B \right), \tag{8}$$

where we used the relations $\text{Re}\{\bar{E}\} = \sqrt{\bar{Q}}\cos\bar{\varphi}$ and $\text{Im}\{\bar{E}\} = \sqrt{\bar{Q}}\sin\bar{\varphi}$. As above, the bar indicates averaging over short time intervals and shows that $\bar{Q}$ and $\bar{\varphi}$ should be found from the corresponding rate equations without stochastic terms. Note that averaging in the last term of Eq. (8) does not follow directly from the Itô formula given by Eq. (7). We introduced it *ad hoc* in order to exclude fluctuations from intensities of Wiener processes. Differential for the phase can be, in turn, obtained using the relation $d\varphi = \text{Im}\{E^* dE\}/Q$ [1]. Substituting Eq. (1) into this relation and using Eq. (3) we obtain

$$d\varphi = \frac{\alpha}{2\tau_{ph}}(G_L - 1)dt + \sqrt{\frac{C_{sp}\bar{R}_{sp}}{2\bar{Q}}} \left( \cos\bar{\varphi} dW^B - \sin\bar{\varphi} dW^A \right). \tag{9}$$

Thus, Langevin forces for $Q$ and $\varphi$ are explicitly defined by

$$F_Q dt = 2\sqrt{\frac{C_{sp}\bar{R}_{sp}\bar{Q}}{2}} \left( \cos\bar{\varphi} dW^A + \sin\bar{\varphi} dW^B \right),$$
$$F_\varphi dt = \sqrt{\frac{C_{sp}\bar{R}_{sp}}{2\bar{Q}}} \left( \cos\bar{\varphi} dW^B - \sin\bar{\varphi} dW^A \right). \tag{10}$$

Corresponding diffusion coefficients given by the relation $\langle F_j(t)F_k(t-\tau) \rangle = 2D_{jk}\delta(\tau)$ are as follows:

$$D_{QQ} = C_{sp}\bar{R}_{sp}\bar{Q}, \quad D_{\varphi\varphi} = \frac{C_{sp}\bar{R}_{sp}}{4\bar{Q}}, \quad D_{Q\varphi} = 0, \tag{11}$$

in agreement with derivation of C. Henry [3].

Equation for the currier number in the differential form can be written as follows [1, 2]:

$$dN = \frac{I}{e}dt - \frac{N}{\tau_e}dt - \frac{QG_L}{\Gamma\tau_{ph}}dt + F_N dt, \tag{12}$$

where $I$ is the pump current, $e$ is the absolute value of the electron charge, and $F_N$ is a real Langevin force driving carrier fluctuations. It is well-known that fluctuations of $N$ are uncorrelated with fluctuations of the phase $\varphi$, but are cross-correlated with fluctuations of $N$ [5]. Corresponding diffusion coefficients are [5]:

$$D_{NQ} = -D_{QQ}, \quad D_{N\varphi} = 0. \tag{13}$$

The diffusion coefficient corresponding to autocorrelations of $F_N$ can be found by considering fluctuations of $N$ as a shot noise. In this case, $2D_{NN}$ is just the sum of rates in and rates out [5]. From the rate equation for $N$ (or, if write explicitly, for $\bar{N}$) we thus find

$$2D_{NN} = \frac{I}{e} + \frac{\bar{N}}{\tau_e} + \bar{Q}\left( C_{sp}\bar{R}_{sp} + \bar{A} \right), \tag{14}$$

where the gain $G$ was presented as the difference between the emission and absorption: $G_L/(\Gamma\tau_{ph}) = C_{sp}\bar{R}_{sp} - \bar{A}$. Using the fact that $\bar{N}$ can be considered constant during the fluctuation averaging time, we can put $d\bar{N}/dt = 0$ for this short time period. Using the expansion of the gain $G$ on the emission and absorption terms, we then obtain $I/e + \bar{Q}\bar{A} = \bar{N}/\tau_e + \bar{Q}C_{sp}R_{sp}$. Substituting this equality into Eq. (14) we finally have:

$$D_{NN} = \frac{\bar{N}}{\tau_e} + C_{sp}\bar{R}_{sp}\bar{Q}. \tag{15}$$

To find an explicit form of the Langevin force $F_N$, we will use the following method. Let us define $F_N$ as the sum of two independent terms: $F_N = F_{NN} + F_{NQ}$, where

$$\begin{aligned} 2D_{NN} &= \langle F_N(t)F_N(t-\tau)\rangle \\ &= \langle F_{NN}(t)F_{NN}(t-\tau)\rangle + \langle F_{NQ}(t)F_{NQ}(t-\tau)\rangle \end{aligned} \tag{16}$$

and $F_{NQ}$ is chosen such that $\langle F_Q(t)F_{NQ}(t-\tau)\rangle = 0$. From the other hand, we can write:

$$\langle F_N(t)F_Q(t-\tau)\rangle = \langle F_{NN}(t)F_Q(t-\tau)\rangle, \tag{17}$$

whence we obtain using Eq. (13): $F_{NN} = -F_Q$. Substituting this result into Eq. (16) and using the relation $2D_{NN} - 2D_{QQ} = 2\bar{N}/\tau_e$ we have $\langle F_{NQ}(t)F_{NQ}(t-\tau)\rangle = 2\bar{N}/\tau_e\,\delta(\tau)$, which yields

$$F_{NQ}dt = \sqrt{\frac{2\bar{N}}{\tau_e}}dW^C, \tag{18}$$

where we introduced the third Wiener process $W^C$, which is independent of $W^A$ and $W^B$. As a result, we have:

$$\begin{aligned} F_N dt &= -2\sqrt{\frac{C_{sp}\bar{R}_{sp}\bar{Q}}{2}}\left(\cos\bar{\varphi}dW^A + \sin\bar{\varphi}dW^B\right) \\ &+ \sqrt{\frac{2\bar{N}}{\tau_e}}dW^C. \end{aligned} \tag{19}$$

The system of SDEs (8), (9) and (12) with stochastic terms given by Eqs. (10) and (19) can be solved numerically with the Euler-Maryama method [4], which is the simplest time discrete approximation used for integration of SDEs. In this method, solution of the vector valued SDE, $X$, is approximated by a continuous time stochastic process $Y(t)$ satisfying the iterative scheme, which for the $k$-th component has the form:

$$Y_{n+1}^k = Y_n^k + a^k\Delta + \sum_{j=1}^m b^{k,j}\Delta W^j, \tag{20}$$

with the initial value $Y_0^k = X_{t_0}^k = X_0^k$, and where $\Delta = t_{n+1} - t_n$ is the time increment, $Y_n^k \equiv Y^k(t_n)$, and $\Delta W^j = W_{t_{n+1}}^j - W_{t_n}^j$ is a normally distributed (with zero mean value and variance equal to $\Delta$) discrete random increment of the $j$-th component of the multidimensional Wiener process. We can write $\Delta W^j = \xi^j\sqrt{\Delta}$, where $\xi$ is a random variable with standard normal distribution. Finally, let us write our system of SDEs in the form ready for integration:

$$\begin{aligned} N_{n+1} &= N_n + \frac{I_n}{e}\Delta - \frac{N_n}{\tau_e}\Delta - \frac{Q_n}{\Gamma\tau_{ph}}\frac{N_n - N_{tr}}{N_{th} - N_{tr}}(1-\chi_Q Q_n)\Delta \\ &\quad -2\sqrt{\frac{C_{sp}\bar{N}_n\bar{Q}_n}{2\tau_e}}\left(\xi_n^A\cos\bar{\varphi}_n + \xi_n^B\sin\bar{\varphi}_n\right)\sqrt{\Delta} + \sqrt{\frac{2\bar{N}_n}{\tau_e}}\xi_n^C\sqrt{\Delta}, \\ Q_{n+1} &= Q_n + \left(\frac{N_n - N_{tr}}{N_{th} - N_{tr}}(1-\chi_Q Q_n) - 1\right)\frac{Q_n}{\tau_{ph}}\Delta + C_{sp}\frac{N_n}{\tau_e} \\ &\quad +2\sqrt{\frac{C_{sp}\bar{N}_n\bar{Q}_n}{2\tau_e}}\left(\xi_n^A\cos\bar{\varphi}_n + \xi_n^B\sin\bar{\varphi}_n\right)\sqrt{\Delta}, \\ \varphi_{n+1} &= \varphi_n + \frac{\alpha}{2\tau_{ph}}\left(\frac{N_n - N_{tr}}{N_{th} - N_{tr}} - 1\right)\Delta \\ &\quad + \sqrt{\frac{C_{sp}\bar{N}_n}{2\tau_e\bar{Q}_n}}\left(\xi_n^B\cos\bar{\varphi}_n - \xi_n^A\sin\bar{\varphi}_n\right)\sqrt{\Delta}, \end{aligned} \tag{21}$$

where $\bar{N}$, $\bar{Q}$, and $\bar{\varphi}$ can be found from the same system of rate equations without stochastic terms and where we included the gain saturation via substituting $G_L$ by $G = G_L(1-\chi_Q Q)$ in equations for $N$ and $Q$ [1] ($\chi_Q$ is the dimensionless gain compression factor).

## 2. Analytical description of the phase diffusion above threshold

According to C. Henry [6], the phase diffusion variance at steady state above threshold can be written as follows:

$$\langle \Delta\varphi^2(t) \rangle = \frac{C_{sp}N}{2Q\tau_e}\left\{(1+\alpha^2)t - \alpha^2 \frac{e^{-\gamma_d t}\cos(\omega_r t - 3\delta) - \cos(3\delta)}{2\gamma_d \cos(\delta)}\right\}, \quad (22)$$

where $\gamma_d$ and $\omega_r$ are the damping rate and the angular frequency of relaxation oscillations, respectively, and $\cos(\delta) = \omega_r/\sqrt{\omega_r^2 + \gamma_d^2}$. One can show using the small-signal analysis that $\omega_r$ can be approximated by the following relation [1]:

$$\omega_r \approx \sqrt{\frac{I - I_{th}}{I_{th} - I_{tr}} \frac{1}{\tau_{ph}\tau_e}}, \quad (23)$$

where we put $I_{th} = N_{th}e/\tau_e$ and $I_{tr} = N_{tr}e/\tau_e$ for the threshold current and for the current at transparency, respectively. The damping rate, in turn, can be written as follows [6]:

$$\gamma_d = \frac{1}{2}\left(\frac{1}{\tau_e} + \frac{dG}{dN}\frac{Q}{\tau_{ph}} + \frac{C_{sp}N}{\tau_e Q} - \frac{dG}{dQ}\frac{Q}{\tau_{ph}}\right). \quad (24)$$

Using the normalized gain in the form $G = G_L(1 - \chi_Q Q)$ we can write:

$$\frac{dG}{dN} = \frac{1}{N_{th} - N_{tr}}, \quad \frac{dG}{dQ} = -G_L \chi_Q \approx -\chi_Q. \quad (25)$$

Finally, we can write for the carrier and the photon number at steady state:

$$N \approx N_{th}, \quad Q \approx \frac{1}{e}(I_b + I_p - I_{th})\Gamma\tau_{ph}. \quad (26)$$

Combining Eqs. (22)-(26), we get the desired analytical expression of the phase diffusion r.m.s. $\sigma_\varphi = \sqrt{\langle \Delta\varphi^2(t) \rangle}$, which provides the dependence of $\sigma_\varphi$ on the bias current $I_b$ and the gain compression factor $\chi = \chi_Q(2\Gamma\tau_{ph}/\varepsilon\hbar\omega_0)$.

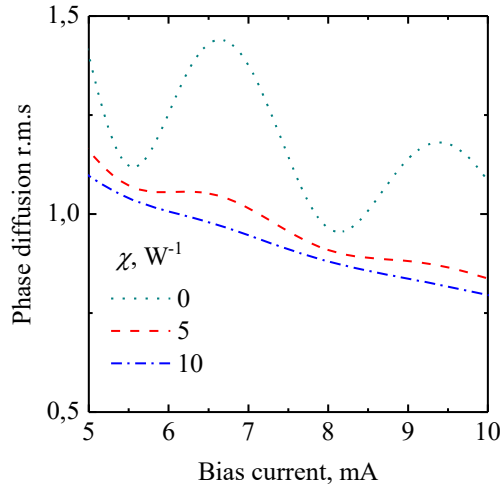

Fig. S1. Phase diffusion r.m.s. as a function of the pump current (or rather the bias current $I_b = I - I_p$) at three different values of the gain compression factor $\chi$.

Figure S1 demonstrates the three $\sigma_\varphi$ curves according to Eq. (22) at different values of $\chi$ for $t = 400$ ps. One can see, that the evolution of the curves is similar to that shown in Fig. 2 in the main text. To make this correspondence more pronounced we plotted $\sigma_\varphi$ against the "bias current", which was just the difference between the pump current $I$ and some fixed value, which we put to $I_p = 12$ mA to be consistent with the main text. The following set of laser parameters was used to draw the curves: $\alpha = 6$, $I_{th} = 10.4$ mA, $I_{tr} = 9.6$ mA, $I_p = 12$ mA, $\Gamma = 0.12$, $C_{sp} = 10^{-5}$, $\tau_{ph} = 1$ ps, $\tau_e = 1$ ns, $\varepsilon = 0.3$, $\omega_0/2\pi = 193.548$ THz.